\documentstyle[epsf]{mn}
\def\epsfsize#1#2{\hsize}

\def\alwaysmath#1{\ifmmode{#1}\else{$#1$}\fi}

\def\etal{{et al.~}}

\def\ers{\alwaysmath{{\rm \, erg\,sec^{-1}}}}
\newcommand\hst{{\it HST}}
\newcommand\rosat{{\it ROSAT}}

\begin{document}

 \title[Faint Ultraviolet Objects in the Core of M13]
 {Faint Ultraviolet Objects in the Core of M13: Optical Counterparts of
the Low Luminosity X-ray Source?}

\author[F. R. Ferraro, \etal]
{Francesco~R.~Ferraro$^{1},$   Barbara~Paltrinieri$^{1},$  
Flavio~Fusi~Pecci$^{1,2},$  
\newauthor
Robert~T.~Rood$^{3},$ and Ben~Dorman$^{3,4}$  \\
$^{1}$Osservatorio Astronomico di Bologna, via Zamboni 33, I-40126
Bologna, ITALY \\
$^{2}$Stazione Astronomica di Cagliari, 09012 Capoterra, ITALY \\
$^{3}$Astronomy Dept, University of Virginia,
	P.O.Box 3818, Charlottesville, VA 22903-0818 \\
$^{4}$Laboratory for Astronomy \& Solar Physics, 
Code 681, NASA/GSFC,	Greenbelt MD 20771
}

\maketitle

\begin{abstract}
The core of the galactic globular cluster M13 (NGC~6205) has been
observed with WFPC2 on the Hubble Space Telescope through visual, blue
and mid- and far-UV filters in a programme devoted to study the UV
population in a sample of Galactic globular clusters.  In the UV Color
Magnitude Diagrams derived from the \hst\ images we have discovered
three faint objects with a strong UV excess, which lie significantly
outside the main loci defined by more than 12,000 {normal} cluster
stars.  The positions of two of the UV stars are nearly coincident
(7\arcsec\ \& 1\arcsec) to those of a low luminosity X-ray source
recently found in the core of M13 and to a 3.5$\sigma$ peak in the
X-ray contour map. We suggest that the UV stars are physically
connected to the  X-ray emission. The UV stars are very
similar to the quiescent nova in the globular cluster M80, and they
might be a, perhaps new, subclass of cataclysmic variable.

\end{abstract}


\section{Introduction} A growing number of Galactic Globular Clusters
(GGCs) have been found to contain X-ray sources.  Hertz and Grindlay
\shortcite{hg83} first proposed that two distinct populations of X-ray
sources might exist in globular clusters: high luminosity X-ray sources
with $ L_{X} > 10^{34.5} \ers $ and low luminosity X-ray sources with $L_X
< 10^{34.5} \ers$ (hereafter LLGCXs). There are about a dozen high
luminosity X-ray sources found in early surveys, and now more than 30
LLGCXs sources in 19 GCCs are listed in the recent compilation of Johnston and
Verbunt \shortcite{jv96}. The presence of X-ray bursts provides compelling
evidence that the high luminosity objects are binary systems with an
accreting neutron star (the so-called Low Mass X-ray Binaries, LMXB).  The
nature of LLGCXs is still not clear.  Recently, Bailyn (1995, but see also
Verbunt \etal\ 1994 and Hasinger, Johnston,
\& Verbunt 1994 for a full discussion of the topic) has reviewed the
most plausible formation scenarios for LLGCXs and has concluded that
the brightest LLGCXs (with $L_X > 10^{32} \ers$) might be transient
neutron star binary systems in quiescent state (since no X-ray burst
has been identified in these objects). Alternatively, they might
result from the superposition of multiple faint LLGCXs as recently
found in NGC~6397 and NGC~6752 (Cool
\etal\ 1993, 1995).  These fainter LLGCXs ($L_X < 10^{32} \ers$) might
be cataclysmic variables (CVs), binary systems in which a white dwarf
(instead of a neutron star) is accreting material from a late type dwarf,
i.e. a MS-SGB star \cite{para83,hg83}. 
Some of the dimmest LLGCXs do appear to be CVs (Cool \etal\ 1993, 1995).
But in other clusters like 47~Tuc, there are other objects in addition to the
CVs found by Paresce, deMarchi, \&
Ferraro (1992) that could be the source of the observed X-rays
\cite{hasinger94}. Most of the CVs located in core-collapsed clusters
could have been created by dynamical processes
\cite{hv83,dr94,bailyn95}, while the CVs in low-density clusters can
be primordial binary systems \cite{vm88}.

The search for optical counterparts for LLGCXs is essential to
determining their origin and the role of the
dynamical history of the parent cluster.  The advent of the Hubble
Space Telescope (\hst) has made it possible to search for candidate
objects in the most crowded regions of the GGCs. Still, some additional
identifying characteristic must be used to select the optical
counterpart of a LLGCX from the 100's of stars within the typical
X-ray position error box. For example, blue, variable objects
(candidate CVs) have been found to fall within the error boxes of some
LLGCXs (Paresce \etal\ 1992, Cool \etal\ 1993, 1995). However, Cool
\shortcite{cool97} shows that colour alone, particularly if based on
$V$ and $I,$ may not select CVs.  Here we present the possible
identification (based on \hst-WFPC2 visual and UV images) of the
optical counterpart of a LLGCX recently discovered in the core of M13
\cite{f96}.

\section{Observations}

 \hst-WFPC2 frames were obtained on January 1996 (Cycle 5: GO 5903
PI: F. R. Fer\-rar\-o) through $V$ (F555W), 
$B$ (F439W), 
$U$ (F336W), mid-UV (F255W)
and UV (F160BW).  The exposures in each filter are listed in Table~1.
A detailed description of the data processing
will be presented in a forthcoming paper \cite{data97}. Here we
summarize just the main steps:

\begin{enumerate}
\renewcommand{\theenumi}{(\arabic{enumi})}

\item All of  the exposures were aligned using MIDAS, and then 
combined to yield a master median frame. 

\item The search for objects was performed on the master 
median frame using ROMAFOT \cite{roma}.

\item PSF-fitting was then performed on each individual frame,
and the instrumental magnitudes were then averaged. In the case of the 
F160BW frames, we used aperture photometry instead of the PSF-fitting
technique because of the high aberration affecting the 
stellar images.

\item All of the instrumental magnitudes have been converted to a fixed 
aperture photometry and then calibrated to the Johnson system using
equation 8 and Table 7 in Holtzmann \etal \shortcite{holt1995}. F160BW
and F255W magnitudes have been calibrated to STMAG system using table
9 by Holtzmann \etal \shortcite{holt1995}.

\end{enumerate}

\begin{table}
\centering
\label{expos}
\caption{Filter and exposure time of the observation}
\begin{tabular}{@{}lrrr}
\hst\ Frames: & $F160BW$ & $4 \times ~300\,$sec  \\ 
 & $F255W$ & $4 \times 50\,$sec  \\
 & $F336W$ & $4 \times 140\,$sec  \\
  & $F555W$ & $4 \times 8\,$sec & $4 \times 1\,$sec  \\ 
\end{tabular}
\end{table}

We adopt distance modulus $(m-M)_0=14.29$ \cite{d93}, a
reddening $E(B-V)=0.02$ \cite{p93}, and the extinction law of
Savage and Mathis \shortcite{sm79}.

\section{Results}

Fox \etal \shortcite{f96} presented \rosat\ High Resolution Imager (HRI)
observations of the GGC M13 and identified 12 X-ray
sources in the field of view of M13 (eight of these had been
previously identified in \rosat\  Position Sensitive
Proportional Counter [PSPC] observations by Margon \etal
1994). Because faint X-ray sources are often associated with distant
quasars, Fox \etal \shortcite{f96} searched the catalogue of Hewitt $\&$
Burbidge \shortcite{hb93} and found that no known quasars fall within
the M13 field of view. However, they were unable to associate most of
the sources to any optical counterpart due to crowding in the optical
images at their disposal. However, their source G (hereafter M13X-G)
was only $\sim 37\arcsec$ from the cluster center
(assumed at $\alpha$ $_{2000}$=16 41 41.5, $\delta$ $_{2000}$=36 27
37.0, Trager, King, \& Djorgovski 1995), so they concluded it was a
LLGCX with a cluster membership probability of $\sim 96\%$.

The  location of  M13X-G is $\sim 30$\arcsec\ E
and 22\arcsec\ N from the cluster center, which is
within the field of view of the WF2 chip in our data.
For this reason we first reduced the WF2 data set and carefully searched 
for any object with markedly blue excess. 

\begin{figure}
\epsffile{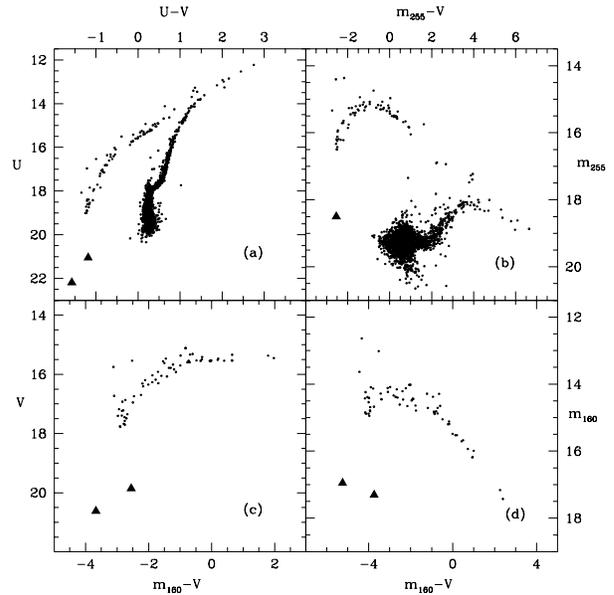}
\caption{
\protect\label{cmds}
$UV$ color magnitude diagrams for M13.
Only stars detected in the WF2 chip have been plotted.
The two faint UV stars  \#23081 (lower)  and \#21429
(upper) that appear to be close to the Fox \etal\ 1996
detections are plotted as filled triangles.}
\end{figure}

Fig.~\ref{cmds} shows the CMDs of stars detected using the WF2 data, in 4
different planes: ($V,~U-V$), ($m_{255},~m_{255}-V$), ($U,~
m_{160}-U$), ($m_{160},~m_{160}-V$).  In the frames taken
through the F160BW filter the main sequence and red giant branch
are invisible, and only the UV bright horizontal branch (HB) stars and
a few fainter stars, including our candidates, are detected.

Inspection of these diagrams shows two extremely blue, low luminosity
objects, namely $\#21429$ and $\#23081$ in our list, which stand
significantly outside the main loci defined by the other cluster
stars.  The positions and magnitudes of these objects are
listed in Table~2. These objects are close to our
photometric detection threshold in all filters, and were not detected
in the $B$ band because the exposures were too short.

\begin{table*}
 \label{cands}
 \centering
 \caption{M13 UV Bright Stars and X-ray Objects}
 \begin{tabular}{@{}lllccccr}
 & $\alpha_{2000}$ & $\delta_{2000}$&  $V$ & $U$ & $m_{255}$ & $m_{160}$ & $\log\,L_x$ \\
Star 23081: & 16 41 43.68 & 36 28 04.67 & 22.192 & 20.616 & .... & 16.948 & \\
M13X-G: & 16 41 44.0 & 36 27 59 & & & & & 32.81 \\
Star 21429: & 16 41 45.1 & 36 27 39 & 21.046 & 19.859 & 18.503 & 17.306  \\
M13X-G-SE: & 16 41 45.1 & 36 27 38 & & & & & $\sim 32.47$ \\
Star 31930: & 16 41 43.90 & 36 26 49.78 & 21.508 & 19.853 & 18.773 & 16.818\\
\end{tabular}
\end{table*}

The possible variability of the two sources was examined by analysing
each frame separately. No clear indication of variability was revealed
from this analysis, but we cannot strongly exclude this possibility
since the $S/N$ is quite low.

\begin{figure}
\epsfxsize=8.0cm
\epsffile{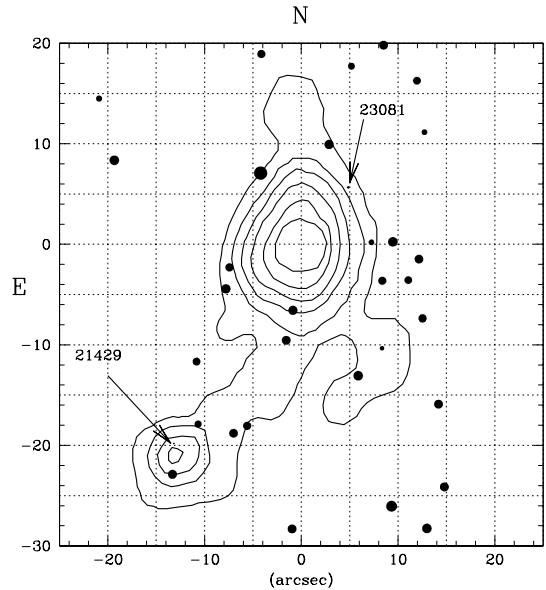}
\caption{
\protect\label{map} Map of the image taken through F160BW image in the
region of the X-ray source M13X-G [located at (0,0)].  The $x$ and $y$
scales are in arcsec. The X-ray emission contour map has been overplotted.
The location of the two UV-stars discovered in this study are indicated by
arrows.}
\end{figure}

Fox \etal \shortcite{f96} reported the absolute position of the M13X-G
source at $\alpha_{2000}$=16 41 44.0, $\delta_{2000}$=36 27 59.0.
Figure~\ref{map} shows a region of $\sim 60\arcsec \times 60\arcsec$
centered on the nominal position of M13X-G. The contours of the
X-ray emission from \rosat\ image retrived from the archives have been
overplotted on a numerical map of the F160BW image.  The locations of the
two UV objects are indicated by arrows.  The position of star $\#23081$ is
$\sim 7$\arcsec\ NW of the main bump of the X-ray map. Star $\#21429$ is
about 1\arcsec\ N of a 3.5$\sigma$ bump that Fox \etal \shortcite{f96}
interpreted as a background fluctuation. We suggest that the
``fluctuation'' is real and will refer to it as M13X-G-SE. Its approximate
position and flux were measured from the archival image and are given in
Table~2. Because of the small number of counts the position is
much less well determined than that of M13X-G. While \rosat\ coordinates are
generally good to ($<5$\arcsec), errors of 10\arcsec\ in absolute position
are not uncommon in HRI images \cite{davidetal92}.  In our view, the
connection between the UV stars and the X-ray emission detected in the
core of M13 is strong. Among the more than 3,000 measured stars in the WF2 region
(80 $\times$ 80 arcsec), we found only 2 UV objects lying outside the
CMD main branches. There are only 3 faint UV objects in the entire
 \hst\ field of
view containing more than 12,000 normal stars.  Only about 130 stars
are within 7\arcsec\ of the X-ray source, so the probability of a
chance coincidence of one UV object with M13X-G is only about 3\%.
The case would be greatly strengthened if deeper observations confirm the
reality of M13X-G-SE.

In subsequent data analysis we have found a third UV object (\#31930) in WF3, 
but none in WF4 and the PC. The archival X-ray image shows no excess of
emission at the location of the third UV star. However, this UV star might
also be associated with a LLGCX since LLGCXs are known to be highly 
time-variable
\cite{hasinger94,hgb93}. Indeed, we predict that further \rosat\
observations of M13 may find a LLGCX at this position.

In addition to being associated with LLGCXs, we think it is likely
that the UV stars are also CVs. Perhaps, the most solid indication for
this comes from Shara \& Drissen (1995). They have recently found a UV
object in the globular cluster M80, which they associate with T~Sco, a
nova observed in 1860.  The near UV properties of T~Sco are very
similar to those of our objects. There is possibly a LLGCX associated
with T~Sco, but the position from the \rosat\ PSPC is not accurate
enough for a definitive identification \cite{hcjv97}. The derived
X-ray to optical flux is also compatible with the value ($f_V/f_X \sim
0.8-4.2$) found by Cool \etal (1995) for three candidate CVs in NGC
6397. The UV to blue ratio of our objects is also similar to that of
star~1 in Sosin
\& Cool (1995), a possible CV/LLGCX in NGC~6624.

The connection to field CVs is more equivocal.  From the photometric
data listed in Table~\ref{cands}, we find $M_V=7.8$ and $M_V=6.7$ for
$\#23081$ and $\#21429$, respectively.  These figures are in good
agreement with the typical absolute magnitude for CVs in the field
($<M_V> \sim +7$, for the dwarf novae CVs; see van Paradijs 1983). The
X-luminosity of M13X-G is $L_X \sim 6.5 \times 10^{32} \ers$, in the
(0.1--2.4\,keV) band, quite compatible to that of field CVs
($L_X<10^{33} \ers$) in the 0.2--4\,keV band
\cite{patterson85}.  
The ratio of X-ray to visible is ${{L_X(0.1-2.4\,{\rm keV})} / {L_V}}
\sim 1$, which is well within the range covered by the field CVs (see
Fig. 3 in Patterson \& Raymond 1985). We can make a similar comparison
using only the softer X-rays observed by \rosat\ by using Fig.~1 of
Hakala \etal (1997). They plot the PSPC counting rate in channels 11-235
against $V$. The 58 HRI counts of Fox \etal\ (1996) correspond to
about 200 PSPC counts and a counting rate of about $10^{-2}\,{\rm
ctn\,sec^{-1}}$. This places it well above band occupied by CVs, but
very near T~Sco.  

The observed flux distribution also bears on this question. One can 
characterize the UV radiation in terms of a blackbody
equivalent even if the radiation is not blackbody. Doing so we find a
lower limit on the temperature of these stars of $T_{\rm BB} >
40,000\,$K, within the temperature range
(10--60000\,K) estimated for 15 field CVs by Patterson \& Raymond
\shortcite{patterson85}.

\begin{figure}
\epsfxsize=8.0cm
\epsffile{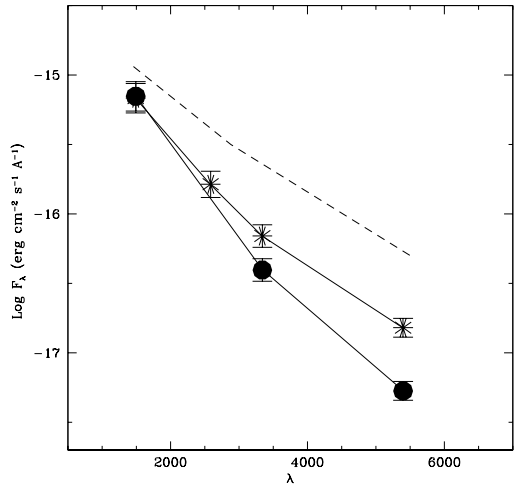} 
\caption{
\protect\label{flux}
Fluxes of star 23081 (large filled circles) and
21429 (big asterisks) in the four passbands used.
The dashed line represents the  flux distribution of the dwarf novae U Gem,
scaled to the distance of M13,
from Figure 2 by Verbunt (1987).
}
\end{figure}

In order to make a more quantitative comparison with field CVs, in
Fig.~\ref{flux}  we plot the flux of the two stars as a function of
wavelength.
The data for the two stars have been plotted as big filled circles
($\#23081$) and big asterisks ($\#21429$), respectively.  The flux
distribution of the dwarf novae U~Gem, scaled to the distance of M13,
has been also plotted for comparison (from Figure 2 by Verbunt 1987).
Verbunt (1987) compiled the UV properties of field cataclysmic
variables based on IUE spectra. He used the parameter $F=\log
f_{1460} - \log f_{2880}$ (where $f_i$ is the flux in ${\rm
erg\,cm^{-2}\,\AA^{-1}}$ at the $i$ wavelength) to characterize the
slope of the ultraviolet continuum flux distribution.  From the data
plotted in Fig.~\ref{flux} we find $F=0.96$ and $0.76$ for star \#23081
and \#21429, respectively.  These values are higher than the mean values
for the field CVs. The value for \#21429 is consistent with the steepest
IUE CV spectra (see figure 3 by Verbunt 1987), and while the value for
\#23081 is higher than that observed for any of Verbunt's sample, it does
not seem out of the question.

Thus, while the cluster UV objects and their possible X-ray counterparts
have some similarities to the field CVs they are not exactly the same.
In particular, the slope of their UV flux distribution may be higher
and the X-ray luminosity may be high relative to the visible.  Because
they are older and they live in a dramatically different environment,
it seems reasonable that cluster objects might be a new class with
properties that differ from those in the field.

\section{Conclusion}

\hst-WFPC2 images of the core of M13 have been used to search for candidate
optical counterparts to the Low Luminosity X-ray source recently found 
in the core of M13 (M13X-G) \cite{f96}.

Based on UV-CMDs we have discovered two extremely blue objects that
lie well outside the main sequences and are spatially very close to
the X-ray source. One UV-star is $\sim 7\arcsec$ from the nominal
position of the X-ray source.  The second UV bright star is $\sim
1$\arcsec\ from a 3.5$\sigma$ X-ray contour map bump, which we
suggest is a weak X-ray source. A third UV object has no corresponding
X-ray emission. There are no other similar objects in our $>
12,000$ star \hst\ sample.

Our objects are very similar to the quiescent nova T~Sco in the
globular cluster M80 and candidate CVs in other clusters.  If UV
bright stars are physically connected to the detected X-ray emission,
then their optical/X-ray properties are similar to but different from those
of field CVs. If they are CVs they may represent a new subclass.
Our observations lack the sensitivity to detect the variability which
would cement the identification as CVs. Detection of variability
and/or H$\alpha$ emission are obvious goals for future
\hst\ studies.  

Globular cluster CVs are ordinarily suspected of being formed in
stellar interactions.  However, several lines of evidence suggest that
observed CVs and LLGCXs are not formed as predicted by classical two
body stellar capture: Johnson \& Verbunt (1996) note that the number
of LLGCXs per cluster scales much less rapidly with cluster density
than expected.  Shara \& Drissen note that there are far fewer
CVs than expected in dense cluster cores like that of M80. If our
conjectures concerning the UV objects and LLGCXs in M13 are confirmed,
it will lengthen this chain of arguments, since M13 is not a
post-core-collapse cluster and has moderate central density.

RTR \& BD are supported in part by NASA Long Term Space Astrophysics 
Grant NAGW-2596 and STScI/NASA Grant GO-5903. The financial support 
by the {\it Agenzia Spaziale Italiana} (ASI) is gratefully acknowledged.
An anonymous referee provided very detailed and helpful comments.
RTR acknowledges useful conversations with Craig Sarazin, and we
especially thank him for assistance with the \rosat\ archival data.


\begin{thebibliography}{}

\bibitem[\protect\citename{Bailyn }1995]{bailyn95}
Bailyn C.D., 1995, ARA\&A, 33, 133

\bibitem[\protect\citename{Buonanno }1983] {roma} 
Buonanno R., Buscema G., Corsi C.E., Iannicola G., 1983, A\&A, 126, 278

\bibitem[\protect\citename{Cool \etal }%
1993]{cool93}
Cool A.M., Grindlay J.E., Krockenberger M., Bailyn C.D., 1993,
ApJ, 410, L103

\bibitem[\protect\citename{Cool \etal }%
1995]{cool95}
Cool A.M., Grindlay J.E. Cohn H.N., Lugger P.J., Slavin S.D.,
1995, ApJ, 439, 695

\bibitem[\protect\citename{Cool }1997]{cool97}
Cool A. M., 1997, in Rood R., Renzini A., eds,
Advances in Stellar Evolution. CUP, Cambridge, p. 191

\bibitem[\protect\citename{Cordova \& Mason } 1983]{cm83}
Cordova F.A., Mason K.O., 1983, in Lewin W. H. G., Van den Heuvel E.P.J., eds,
Accretion Driven Stellar X-ray Sources. CUP, Cambridge, p. 147

 \bibitem[\protect\citename{David \etal }%
1992]{davidetal92} David L. P., Harnden F.R., Jr., Kearns K. E.,
Zombeck, M. V., 1992,  The \rosat\ High
Resolution Imager, U. S. \rosat\ Science Data Center, p. 4 

\bibitem[\protect\citename{Di Stefano \& Rappaport }1994]{dr94}
Di Stefano R., Rappaport S., 1994, ApJ, 423, 274

\bibitem[\protect\citename{Djorgovski }1993] {d93}
Djorgovski S. G., 1993, in Djorgovski S. G., Meylan G., eds, The
Structure and Dynamics of Globular Clusters. ASP, San Francisco, 
p. 373

\bibitem[\protect\citename{Ferraro et~al. }1997]{data97}
Ferraro F. R., Paltrinieri B., Fusi Pecci F., 
Cacciari C., Dorman, B., Rood, R. T.,
1997, in preparation

\bibitem[\protect\citename{Fox \etal }%
1996]{f96}
Fox D., Lewin W., Margon B.,
van Paradijs J., Verbunt F., 1996, MNRAS, 282, 1027


\bibitem[\protect\citename{Grindlay }%
1994]{g94} Grindlay J.E., 1994 in Fruchter A. S., Tavani M., Backer
D. C., eds, ASP Conf. Ser. 72, Millisecond Pulsars: a Decade of
Surprise. ASP, San Francisco,  p. 57

\bibitem[\protect\citename{Hakala \etal }1997]{hcjv97}
Hakala P. J., Charles P. A., Johnson H. M., Verbunt F., 1997,
MNRAS, 285, 693

\bibitem[\protect\citename{Hasinger \etal }%
 1994]{hasinger94} 
Hasinger G., Johnston H. M., Verbunt F., 1994, A\&A, 288, 466 

\bibitem[\protect\citename{Hertz \& Grindlay }1983]{hg83} 
Hertz P., Grindlay J.E., 1983, ApJ, 267, L83 

\bibitem[\protect\citename{Hertz, Grindlay, \& Bailyn }1993]{hgb93}
Hertz P., Grindlay J. E., Bailyn C. D., 1993, ApJ, 410, L87

\bibitem[\protect\citename{Hewitt \& Burbidge }1993]{hb93}
Hewitt A., Burbidge G., 1993, ApJS, 87, 451

\bibitem[\protect\citename{Holtzmann \etal\ }%
1995]{holt1995}
Holtzmann J. A., Burrows C. J., Casertano S., Hester J. J.,
 Trager J. T., Watson A. M., Worthey G., 1995, PASP, 107, 1065

\bibitem[\protect\citename{Hut \& Verbunt }1983]{hv83}
Hut P., Verbunt F., 1983, Nature, 301, 587

\bibitem[\protect\citename{Johnston \& Verbunt }1996]{jv96}
Johnston H. M., Verbunt F., 1996, A\&A, 312,80

\bibitem[\protect\citename{Margon \etal\ } 1994]{margon94}
Margon B., Deutsh R., Silber A., Lewin W. H. G., van Paradijs J., Van
der Klis, M., 1994, in Makino F., Ohashi T., eds, New Horizon of X-ray
Astronomy. University Academy Press, Tokyo, p. 395

\bibitem[\protect\citename{Paresce \etal\ } 1992]{pdf92}
Paresce F., De Marchi G., Ferraro F.R., 1992, Nature, 360, 46

\bibitem[\protect\citename{Paresce \& De Marchi } 1994a]{pd94a}
Paresce F., De Marchi G., 1994, A\&A, 281, L13

\bibitem[\protect\citename{Paresce \& De Marchi } 1994b]{pd94b}
Paresce F., De Marchi G., 1994, ApJ, 427, L33

\bibitem[\protect\citename{Patterson \& Raymond }1985]{patterson85}
Patterson J., Raymond J., 1985, ApJ, 292, 535

\bibitem[\protect\citename{Peterson }1993] {p93}
Peterson C. J., 1993, in Djorgovski S. G., Meylan G., eds, The
Structure and Dynamics of Globular Clusters. ASP, San Francisco,
p. 337

\bibitem[\protect\citename{Savage \& Mathis }1979] 
{sm79} Savage B. D., Mathis J. S., 1979, ARA\&A, 17, 73

\bibitem[\protect\citename{Shara \& Drissen }1995]{sd95} 
Shara M. M., Drissen L., 1995, ApJ, 448, 203

\bibitem[\protect\citename{Sosin \& Cool }1995]{sc95} 
Sosin C., Cool A. M., 1995, ApJ, 452, L29

\bibitem[\protect\citename{Trager, King, \& Djorgovski }%
1995]{tkd95} Trager S., King I. R., \& Djorgovski S. G., 1995, AJ, 109, 218
 

\bibitem[\protect\citename{van Paradijs }%
1983]{para83} van Paradijs J., 1983, in Lewin W. H. G., Van den
Heuvel E. P. J., eds, Accretion Driven Stellar X-ray Sources.
CUP, Cambridge,  p. 189

\bibitem[\protect\citename{Verbunt }1987]{verbunt87} 
Verbunt F., 1987, A\&AS, 71, 339

\bibitem[\protect\citename{Verbunt \etal\ }1994]{verbunt94} 
Verbunt F., Johnston H. M., Hasinger G., Belloni T., Bunk W., 1994, in
Shafter A. W., ed, ASP Conf Ser. 56, Interacting Binary Stars. ASP, San
Francisco, p. 244

\bibitem[\protect\citename{Verbunt \& Meylan }1988]{vm88}
Verbunt F., Meylan, G., 1988, A\&A, 203, 297 

\end{thebibliography}
\end{document}